\UseRawInputEncoding
\documentclass[12pt,reqno]{article}
\usepackage{amssymb}
\usepackage{amsthm}
\usepackage{amsfonts}
\usepackage{url}
\usepackage{hyperref}
\usepackage{breakurl}
\usepackage{rotating}

\begin{document}
\title{On the existence of critical exponents for self-avoiding walks
}
\author{
Anthony J Guttmann\footnote{ORCID:0000-0003-2209-7192}\\
School of Mathematics and Statistics\\
The University of Melbourne\\
Vic.\ 3010, Australia\\
\href{mailto:guttmann@unimelb.edu.au}{\tt guttmann@unimelb.edu.au}
\and
Iwan Jensen\footnote{ORCID:0000-0001-6618-8470}\\
College of Science and Engineering,\\
Flinders University, Tonsley, GPO Box 2100, \\Adelaide 5001, South Australia\\
\href{mailto:iwan.jensen@flinders.edu.au}{\tt iwan.jensen@flinders.edu.au}
}

\date{}
\maketitle
\begin{abstract}
We describe some ideas of John Hammersley for proving the existence of critical exponents for two-dimensional self-avoiding walks and provide numerical evidence for their correctness.
\end{abstract}

\noindent {\bf AMS Classification scheme numbers}: 05A15, 82B20, 82B27, 82B41

\vskip .2cm


\vskip .3cm

\section{Introduction}
\label{introduction}
The self-avoiding walk was introduced by Orr in 1947 \cite{O47} to model the excluded-volume effect experienced by long-chain polymers in dilute solution. Its metric properties were first considered by Flory \cite{F49} in 1949. Its migration to a problem of interest to mathematicians began with the 1954 paper by Hammersley and Morton \cite{HM54}, and since that time there has been an increasing number of articles reflecting the many areas of science impacted by aspects of this problem. 

In particular, while still of interest to polymer chemists, it has modelled various phenomena in biology, particularly the folding of biological molecules such as DNA, and served as a paradigm of a non-Markovian enumeration problem of interest to computer scientists, algorithm designers and mathematicians working in algebraic combinatorics. 

Many of the recent advances have come about by the application of probability theory, and the self-avoiding walk is a benchmark model of phase transitions to researchers in mathematical physics and statistical mechanics. It has also been applied to problems in  telecommunication networks.

We will consider self-avoiding walks (SAW) on a regular lattice, usually the square lattice ${\mathbb Z^2}$ or another regular two-dimensional lattice such as the hexagonal or triangular lattice. An $n$-step SAW $\omega$ beginning at  site ${\bf x}$ is a sequence of sites $(\omega(0), \omega(1), \ldots , \omega(n))$ with $\omega(0)={\bf x}$ and the distance between adjacent sites $|\omega(i)-\omega(i+1)| = 1.$ The self-avoidance condition implies that $\omega(i) \ne \omega(j)$ for $i \ne j.$ We denote the length of the walk by $n=|\omega|-1$ and the set of all walks of length $n$  by $\Omega_n$.

Let $c_n=|\Omega_n|$ denote the number of $n$-step SAWs distinct up to translation. Thus on the square lattice $c_0=1$ by convention, $c_1=4,$ $c_2=12,$ $c_3=36$ and $c_4=100,$ which is the first manifestation of the restriction imposed by the self-avoiding condition. The generating function is $$C(z)=\sum_{n\ge 0} c_n z^n,$$ and the subset of SAWs whose end-point is adjacent to their starting point are called {\em self-avoiding polygons,} as they are indeed topological polygons if an additional step joining the origin to the end-point is drawn. The number of $n$-step polygons is denoted $p_n$, and their generating function is $$P(z)=\sum_{n\ge 0} p_n z^n.$$ Apart from these two generating functions, certain metric properties are of interest, such as the mean-square end-to-end distance of SAWs $$\langle R^2\rangle_n = \frac{1}{c_n}\sum_{\; \Omega_n} |\omega(n)|^2.$$

If one concatenates two SAWs, say of length $n$ and $m$, so that the origin of one coincides with the end-point of the other, one produces an $n+m$ step walk which may or may not be self-avoiding. Thus $$c_n c_m \ge c_{n+m}.$$ This is a {\em sub-multiplicative} inequality, and taking logarithms produces the corresponding {\em sub-additive} inequality $$\log{c_n} +\log{c_m} \ge \log{c_{m+n}}.$$ Then by Fekete's lemma \cite{F23} one has $$\lim_{n \to \infty} \frac{\log{c_n}}{n} = \inf_{n \ge 1} \frac{\log{c_n}}{n},$$ and the limit $\lim_{n \to \infty}$ exists in $[-\infty,\infty).$ For SAWs in ${\mathbb Z}^n$ it is clear that $\log{c_n} > 1,$ (just consider walks restricted to north and east steps), so the limit, written $\log \mu,$ is finite. Equivalently, one can write $$c_n \ge \mu^n, \,\, n \ge 1,$$ a result first proved by Hammersley \cite{H61}. Hammersley originally called $\log \mu$ the {\em connective constant}, but more recently $\mu$ itself has been referred to as the connective constant by many authors, including Hammersley \cite{H91}.

In fact it is universally believed that $$c_n \sim A \mu^n n^{\gamma-1}$$ where $\gamma$ is called the {\em critical exponent.} The nature of this exponent can also be understood probabilistically. Consider two $n$-step SAWs, $\omega_1$ and $\omega_2$, with a common origin ${\bf 0}$. The probability that they don't intersect is $${\mathbb P}(\omega_1 \cap \omega_2 = \{ {\bf 0}\}) = \frac{c_{2n}}{c_n^2} \sim \frac{D}{n^{\gamma-1}}.$$ For the mean-square end-to-end distance it is expected that $$\langle R^2 \rangle_n \sim B n^{2\nu},$$ where $\nu$ is another critical exponent. Indeed, for two-dimensional SAWs on a regular lattice it is accepted that $\gamma = 43/32,$ and $\nu = 3/4$ exactly \cite{N82}. These results would follow if the scaling limit of SAWs is given by $\textrm{SLE}_{8/3}$ \cite{LSW02}, and again, while this is widely accepted, it has not been proved. 

In three dimensions there is no conjectured exact value, but the best numerical estimate, due to Clisby \cite{C17} is $\gamma = 1.15695300 \pm 0.00000095.$ For $d=4$ (the so-called {\em upper critical dimension}) it is believed that the exponent is 1, but with a logarithmic correction. More precisely $$ c_n \sim F \mu^n (\log{n})^{1/4}.$$ Indeed, for {\em weakly} SAW there are rigorous results for the log
corrections, described in \cite{BBS15} and \cite{BSTW17}. These coincide with the conjectured results for $d=4$ SAWs.

What has been proved is a result due to Hammersley and Welsh \cite{HW62} more than 50 years ago that $$\mu^n \le c_n \le \mu^n e^{\kappa\sqrt{n}},$$ for SAWs on ${\mathbb Z^d}$ with $d \ge 2.$ This was for $d > 2$ improved a year later by Kesten \cite{K63} who showed that the upper bound could be strengthened to $$c_n \le \mu^n \exp \left (\kappa n^{2/(d+2)}\log{n} \right ).$$ Note that the existence of a critical exponent would imply $$c_n \sim  \mu^n e^{O(\log{n})}.$$ In dimensions $d \ge 5,$ Hara and Slade \cite{HS92} showed that SAWs have the same scaling behaviour as simple random walks, so that, in particular, the critical exponent exists and has the exact value $\gamma=1.$ Recently for $d=2$ Duminil-Copin et. al \cite{DCea18} proved that, for infinitely many values of $n,$ though not for all $n,$ $$ c_n \le \mu^n e^{\kappa(n^{1/2-\epsilon})},$$ for an explicit value of $\epsilon.$ Furthermore, for the hexagonal lattice only, and subject to an unproven conjecture about the behaviour of SAWs on the universal cover of the lattice (that the connective constant is the same as that for SAWs on the regular lattice), they prove that there exist positive constants $C_0$ and $C$ such that $$c_n \le C_0n^C\mu^n.$$

For self-avoiding polygons we are in slightly better shape. Note that for $ {\mathbb Z}^d$ the number of polygons of length $n,$ denoted $p_n$ can be non-zero only for $n$ even. Hammersley \cite{H61} proved that $p_{2n} \sim \textrm{const.} \mu_p^{2n},$ where $\mu_p = \mu,$ the connective constant for SAWs on the same lattice. Madras \cite{NM95} proved the result that 
$$p_n \le \frac{C\mu^n}{\sqrt{n}},$$
for polygons on the square lattice (note that $n$ must be even). In \cite{H18}, Hammond  shows that this can be improved to give 
$$p_n \le \frac{C\mu^n}{n^{3/2+o(1)}},$$

In \cite{DGHM16} Duminil-Copin et al. proved that the probability
that a walk of length $n$ ends at a point ${\bf x}$ tends to 0 as $n$ tends to infinity,
uniformly in ${\bf x}.$ Also, when ${\bf x}$ is fixed, with $||{\bf x}|| = 1,$ this probability
decreases faster than $n^{-1/4+\epsilon}$ for any $\epsilon > 0.$ As this probability is equal to $n p_n/c_n,$ it provides a bound on
the probability that a self-avoiding walk is a polygon.

In Theorem 1.2 of \cite{H19}, Hammond improved this bound to $n^{-1/2+\epsilon}$ for walks of sufficient length,
and subsequently, in \cite{H19a} Hammond made two further improvements. For the special case $d=2,$ the bound was improved to $n^{-4/7+\epsilon}$ on a set $n$ of limit supremum density at least 1/1250. Then, assuming the existence of critical exponents for both SAWs and SAPs, the bound can be further improved to $n^{-2/3+\epsilon}.$
Impressive as these bounds are, we remark that they are far from the expected exact value, which, in two dimensions is $n^{-59/32}.$

For the exponent $\nu$ a similar degree of knowledge (or ignorance) prevails. For $d=1,$ trivially $\nu =1.$ For $d=2$ it is widely accepted that $\nu=3/4.$ For $d=3$ one only has a numerical estimate due to Clisby \cite{C10}, which is $\nu = 0.587597 \pm 0.000007.$ For $d=4$ one expects $$\langle R^2\rangle_n \sim Fn (\log {n})^{1/4},$$ while for $d \ge 5$  Hara and Slade \cite{HS92} have proved that $\nu = 1/2,$ and that the scaling limit is given by Brownian motion. 
In \cite{M14}, for $n$-step SAWs in ${\mathbb Z}^d,$ Madras proved that the mean-square end-to-end distance is at least $n^{4/(3d)}$ times a constant, which implies that $\nu \ge 2/(3d),$ assuming it exists.

However, for $d=2,\,\,3,\,\,4$ almost nothing else has been proved. It is not even known that the mean-square displacement grows at least as rapidly as
simple random walks and slower than ballistically. That is to say, $\langle R^2\rangle_n \ge \textrm{const.} n$ has not been proved, nor that $\langle R^2\rangle_n < \textrm{const.} n^{2-\epsilon}.$ In fact Duminil-Copin and Hammond \cite{DCH13} have proved that the walk is sub-ballistic in the sense that there is an exponentially small probability of its having any given positive speed. Alternatively expressed, they prove that $$\lim_{n \to \infty} \langle R^2\rangle_n/n^2 \to 0.$$ Informally this says that $\nu < 1,$ but proving that the exponent is $2 - \epsilon$ remains an open problem.

On May 20th, 1982 one of us (AJG) had a discussion with the late John Hammersley FRS in which we lamented that the ``obvious'' existence of a critical exponent for two- and three-dimensional SAWs had still not been proved. While we are closer, that still remains true today, some 40 years later. In the course of that discussion, three possible approaches to a proof were sketched by Hammersley. We have investigated each of these numerically, and find them to be almost certainly true. 

The first approach requires the definition of a new sub-class of SAWs which we call worms. The simplest manifestation of these are SAWs in ${\mathbb Z}^2$ whose origin and end-point have the same $x$-coordinate. If these can be proved to be super-multiplicative, then it follows that the number of SAWs $$c_n \le C_1\mu^n n^2.$$ We have enumerated worms on both the square-lattice up to length 59 and on the triangular lattice up to length 40, and (numerically, not rigorously) find them to be super-multiplicative. We also propose a scaling argument for their critical exponent, and provide numerical evidence in strong support of the predicted value. For triangular lattice worms we find, numerically, the even stronger result that the coefficients are log-convex. The same appears to be true for the odd and even sub-sequences of square-lattice worms.

The second idea bears a superficial similarity to the recent approach of Duminil-Copin et al. \cite{DCea18} discussed above. One considers a two-layer lattice with SAWs free to move within either layer or vertically on any bond joining the two layers. If a fraction $p$ of the steps are vertical, then the number of SAWs is $$c_n(p) \sim e^{n\kappa(p) +o(n)}.$$ Clearly, $\kappa(0) = \log(\mu).$ If and only if $$\kappa(p) = \kappa +O(-p\log{p}) \,\,\ {\rm as} \,\, p \to 0,$$ then $$c_n = \exp(\kappa n +O(\log{n})).$$ We have generated some data for this two-layer lattice, and  the behaviour of $\kappa(p)$ as $p \to 0$ can be estimated from our data in a reasonably convincing manner.

The third approach is directly at the coefficient level. Consider the ratio of alternate terms, $t_n = c_n/c_{n-2}.$ Then if $$r_n=\frac{t_{n+2}}{t_n} = \frac{c_{n+2}c_{n-2}}{c_n^2} = 1 + O \left ( \frac{1}{n^2} \right),$$ it follows that $$c_n = \exp(\kappa n + O(\log {n} )).$$ Available data provides abundant support for the conjectured behaviour.

In the next three sections we discuss these three approaches, and provide compelling numerical evidence, but alas not a proof, that support the proposed behaviour in each case.

\section{Worms and sub--additivity}
\begin{figure}[ht] 
   \centering
   \includegraphics[width=4.3in]{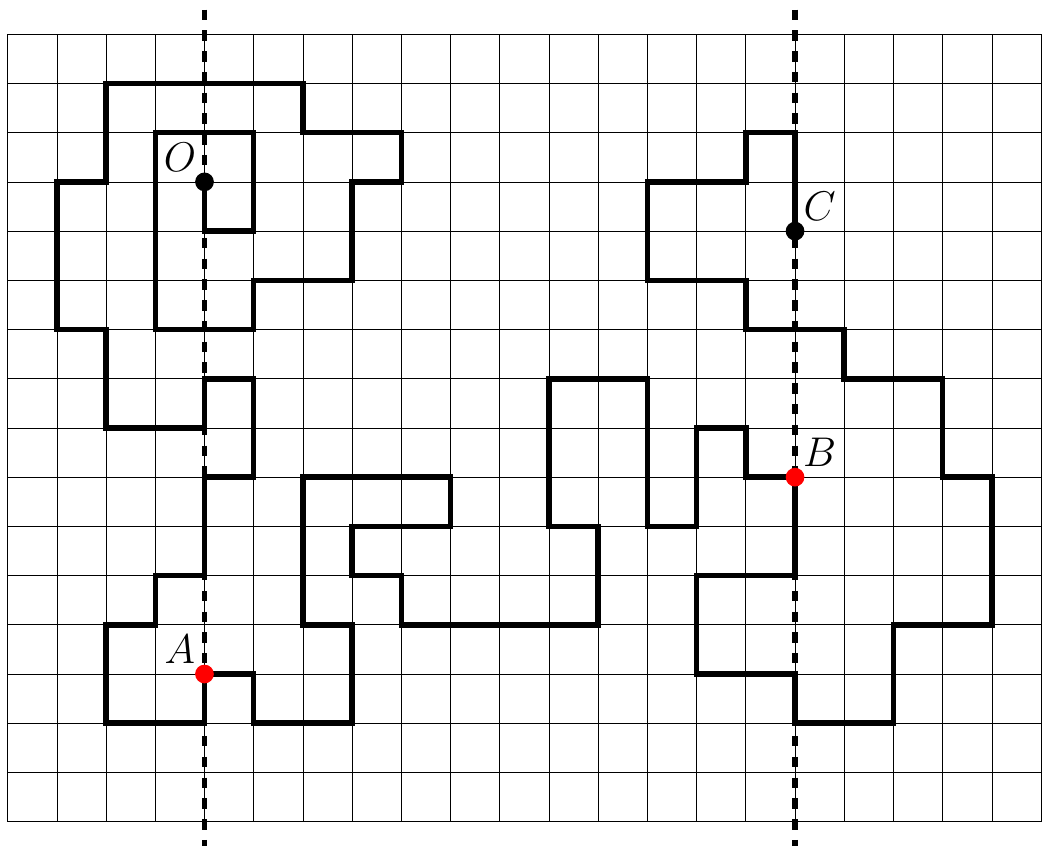} 
 \caption{Schematic showing SAW deconstructed into two worms and a bridge}
   \label{fig:sawcomp} 
\end{figure}

We consider SAWs in ${\mathbb Z}^2.$ A {\em worm} is defined as a SAW whose origin and end-point has the same $x$-coordinate. As shown schematically in Figure~\ref{fig:sawcomp}, any SAW can be subdivided into three parts (one or two of which may be absent in degenerate cases). 
Here $A$ is the last visited vertex of the SAW with the same $x$ coordinate as the origin, and $B$ is the first visited vertex with the same $x$ coordinate as the end-point $C$. Recall that a bridge is a SAW whose origin has unique minimal $x$ coordinate and whose endpoint has (not necessarily unique) maximal $x$ coordinate.
Hence, the walk segment from the origin $O$ to vertex $A$ is a worm, the segment from $A$ to $B$ is a {\em bridge} with a single horizontal step appended to the end (except in the case where the end-points are in adjacent columns and the bridge is just a single step), and the segment from $B$ to the end-point $C$ is another worm. 

Clearly, bridges can be concatenated without possibility of self-intersection, so one has $$b_n b_m \le b_{n+m}.$$ Then by Fekete's lemma one has $$\lim_{n \to \infty} \frac{\log{b_n}}{n} = \sup_{n \ge 1} \frac{\log{b_n}}{n},$$ and the limit $\lim_{n \to \infty}$ exists in $(-\infty,\infty].$ We say that bridges are super-multiplicative.

Let the walk segment $OA,$ which is a worm, be of length $k$. The walk segment $AB$ is a bridge of length $l$, and the segment $BC$ is a worm of length $n-k-l$. It follows from super--multiplicativity of bridges that the number of bridges from $A$ to $B$ is $\le \textrm{const.} \mu^l.$ Then the argument goes that {\em if} worms are also super-multiplicative, then the number of worms from $O$ to $A$ is $\le \textrm{const.} \mu^k,$ and from $B$ to $C$ is $\le \textrm{const.} \mu^{n-k-l}.$

The vertices $A$ and $B$ can be any of the vertices, so there is $n^2$ possible choices for these two vertices. Putting these considerations together, it follows that $$c_n \le \textrm{const.} n^2 \mu^k \mu^l \mu^{n-k-l} = \textrm{const.} n^2 \mu^n.$$

We list in Table \ref{tab:wc} the counts of square-lattice worms of length $< 60$ steps. Numerical experimentation confirms super-multiplicativity for any choice of $n$ and $m$ such that $n+m < 60.$ That the growth constant $\mu$ is the same as that of SAWs follows trivially from the fact that worms are bounded above by the number of SAWs and below by the number of self-avoiding polygons, both of which have the same growth constant. 

We also list the enumeration data for worms on the triangular lattice up to length 40 steps in Table \ref{tab:wc}.

Assuming the existence of an exponent for worms, so that the number of worms behaves as $$w_n \sim F \mu^n n^{\gamma_w-1},$$ we give an argument that $$\gamma_w = \gamma - \nu = 19/32.$$ The argument is extremely simple. The exponent for SAWs is $\gamma.$ The end-point of a SAW is assumed to be radially symmetric, modulo lattice effects. The average length of an $n$-step SAW is $O(n^\nu).$ A typical walk of this length will end on a circle of radius $O(n^\nu)$ centred on the origin, uniformly modulo lattice effects. The circumference is proportional to $n^\nu,$ and so the probability of any particular ray, such as $y=0,$ is proportional to $n^{-\nu.}.$ The number of SAWs is proportional to $\mu^n n^{\gamma-1},$ hence the number of worms is proportional to $\mu^n n^{\gamma-\nu-1}$ 
and so the proportion of SAWs ending on any particular radial line scales as $n^{-\nu}.$ Taking the radial line as the $x$ axis gives the exponent for worms as $\gamma - \nu.$ Hence $\gamma_w=\gamma - \nu.$
An alternative derivation of this scaling law based on the theory of polymer networks \cite{D86} developed by Duplantier was recently given by Duplantier and Guttmann \cite{DG19}.

\begin{figure}[ht] 
   \centering
   \includegraphics[width=4.3in]{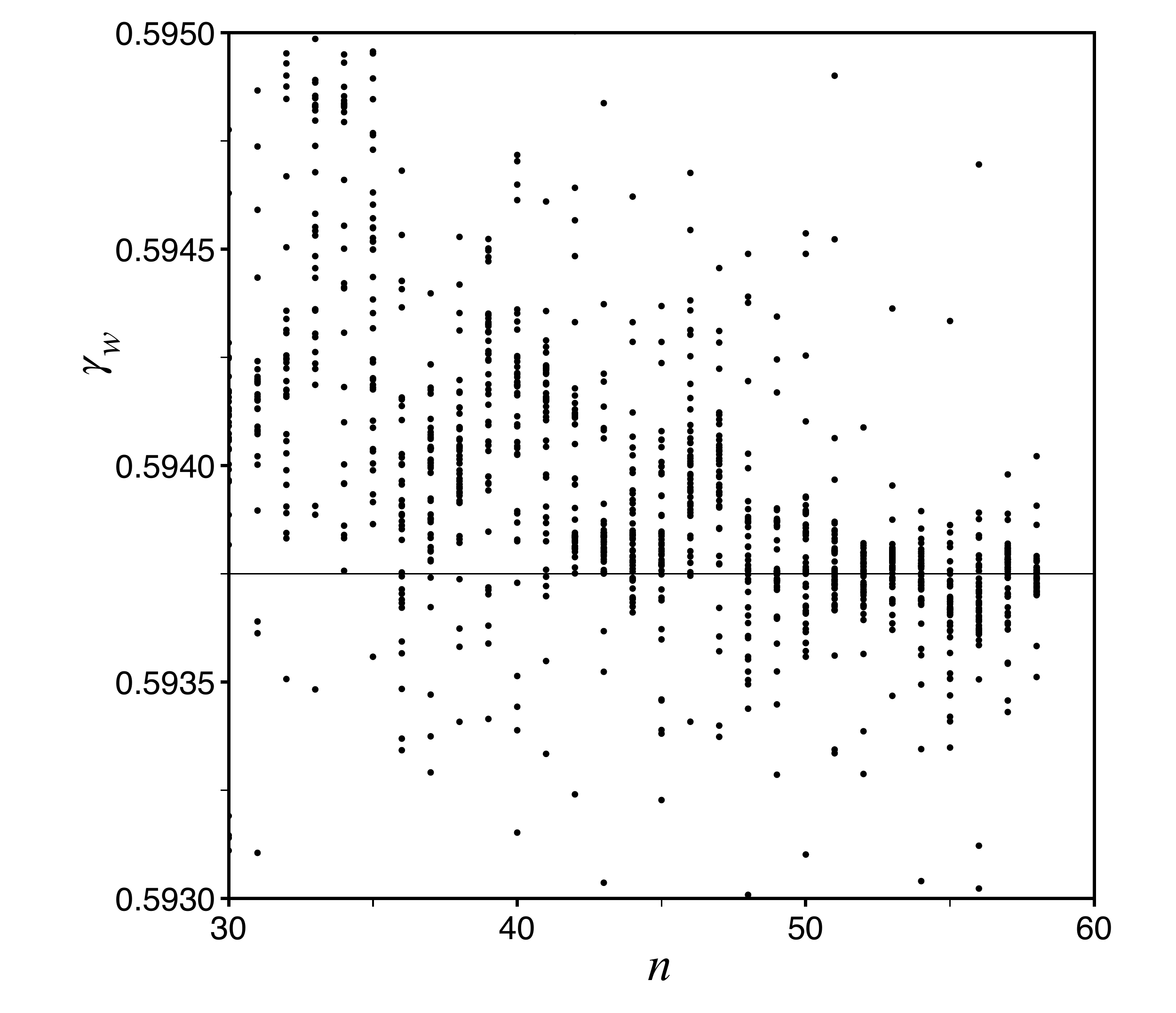} 
\caption{Estimates of the worm exponent $\gamma_w$ for square-lattice worms.}
   \label{fig:gamma}
\end{figure}

We have carried out an analysis of the series in Table \ref{tab:wc} which confirms this result to five or six digits. 
Generating functions for square lattice SAWs tend to have a confluent singularity at the critical point $z_c=1/\mu$, which makes estimating the critical exponent accurately somewhat tricky. Since we have a very accurate estimate for $\mu$ one can overcome this difficulty by using  biased differential approximants as demonstrated in \cite{J16}. In Figure~\ref{fig:gamma} we plot estimates of the exponent $\gamma_w$ obtained from third order biased differential approximants assuming a confluent singularity of order 2. The straight line in the plot indicates the conjectured exact value of $\gamma_w$ and $n$ is the number of terms utilised by the approximants. 

We observe that the triangular lattice worm sequence appears to be not only super-multiplicative but also log-convex, that is to say, the coefficients satisfy $c_{2,k} =c_{k-1}c_{k+1} - c_k^2 \ge 0.$ 
This is a more demanding restriction than super-multiplicativity. Clearly it implies that the ratios of successive coefficients $r_n = c_n/c_{n-1}$ provide an increasing sequence of lower bounds to the growth constant. If this could be proved, we would have the result $\mu \ge 4.1098,$ compared to the best estimate $4.15079\cdots,$ which is an excellent bound. 
Note too that this log-convexity cannot be iterated, that is to say, the sequence $c_{2,k}$ is not log-convex. This is sufficient to prove that the worm sequence is not a Stieltjes moment sequence.

While the square-lattice worm sequence is not log-convex, the subsequence of odd and even terms, treated separately, are (almost) log-convex. There is a failure at the first or second term, but after that the required inequality holds. Again, assuming this persists, one finds the excellent bound on the growth constant, $\mu > 2.6202,$ which is less than $0.7\%$ below the best numerical estimate $\mu = 2.638158553\cdots.$

Our efforts to prove super-multiplicativity of worms have been unsuccessful, but hopefully someone can? We make two remarks that may be helpful in forming a proof. 

Firstly, there is no point in playing with a typical worm as drawn and try and unfold it to prove super-multiplicativity, as the salient feature is not the shape, which can be essentially that of any SAW, but the restriction that the origin and end-point must be on the nominated ray $y=0.$ It is this restriction that induces super-multiplicativity. 

The second comment is that in fact only a subset of worms can be used to construct a SAW in the manner described, as the endpoint must be ``exposed'', to the extent that it can be concatenated with a bridge.
Whether this changes the associated exponent is not clear. It might just change the multiplicative constant $C$ in the expected behaviour $w_n \sim C  \mu^n \cdot n^g.$

Finally, we remark that while super-multiplicativity of worms is sufficient to establish the quoted result, it is not necessary. One could have low-order counts of worms breaking the conditions for super-multiplicativity, yet being asymptotically supermultiplicative, for example, and the result would still hold.

 \begin{table}[htp]
 \scriptsize{
\caption{Number of square and triangular lattice worms $w_n$.}
\centering
\tabcolsep=0.11cm
\begin{tabular}{rrr}
\hline
$n$ & Square & Triangular \\ \hline
 1 & 1 & 1 \\ 
2 & 1 & 3 \\ 
3 & 3 & 11 \\ 
4 & 7 & 41 \\ 
5 & 19 & 155 \\ 
6 & 41 & 603 \\ 
7 & 113 & 2361 \\ 
8 & 261 & 9321 \\ 
9 & 713 & 37015 \\ 
10 & 1681 & 147657 \\ 
11 & 4567 & 591227 \\ 
12 & 10993 & 2374539 \\ 
13 & 29717 & 9561487 \\ 
14 & 72493 & 38585555 \\ 
15 & 195269 & 156007667 \\ 
16 & 481261 & 631806555 \\ 
17 & 1292729 & 2562434223 \\ 
18 & 3211263 & 10405918209 \\ 
19 & 8606801 & 42306525037 \\ 
20 & 21515135 & 172180092143 \\ 
21 & 57561815 & 701397054549 \\ 
22 & 144631085 & 2859651782649 \\ 
23 & 386382359 & 11668050956347 \\ 
24 & 974968645 & 47642140547239 \\ 
25 & 2601469419 & 194655761552949 \\ 
26 & 6587913395 & 795800965884627 \\ 
27 & 17560287513 & 3255243440482761 \\ 
28 & 44605607915 & 13322539042506413 \\ 
29 & 118794020215 & 54550603704403145 \\ 
30 & 302552020141 & 223463822517735377 \\ 
31 & 805154546027 & 915793205476764163 \\ 
32 & 2055349807933 & 3754557467023732917 \\ 
33 & 5466153819177 & 15398587816297681137 \\ 
34 & 13981989972487 & 63176331176917643997 \\ 
35 & 37163553099481 & 259280980073959248551 \\ 
36 & 95232454201457 & 1064441428789266587549 \\ 
37 & 252996576301671 & 4371197941602931769361 \\ 
38 & 649350222217353 & 17955606854760922989315 \\ 
39 & 1724307276343901 & 73775887740799225700335 \\ 
40 & 4432047315550353 & 303206031122586811060823 \\ 
41 & 11764320781845771  \\ 
42 & 30277573156947843   \\ 
43 & 80339307513504149 \\ 
44 & 207011412761661937 \\ 
45 & 549112230724773597  \\ 
46 & 1416425680613326621  \\ 
47 & 3756069524370396305  \\ 
48 & 9698241854071761949  \\ 
49 & 25710902909134456911  \\ 
50 & 66445949659066346623  \\ 
51 & 176111360045162875247  \\ 
52 & 455511686438548203979  \\ 
53 & 1207040286086942540089  \\ 
54 & 3124406504068168856745  \\ 
55 & 8277534048332309786119  \\ 
56 & 21441540272989618597401 \\ 
57 & 56794667899918151597467  \\ 
58 & 147214295603084699698663  \\ 
59 & 389875459357540906081455 \\
\hline
\end{tabular}
\label{tab:wc}}
\end{table}

\section{Walks in a double layer}

In this approach, one considers a two-layer subset of ${\mathbb Z}^3,$ with the walk taking a fraction $p$ of steps in the vertical or $z$ direction. That is to say, there are two square lattices joined by orthogonal edges between vertices in the two lattices. Alternatively, this is a two-layer slice of ${\mathbb Z}^3.$ The number of SAWs is $$c_n(p) \sim e^{n\kappa(p) +o(n)}.$$ Clearly, $\kappa(0) = \kappa=\log(\mu).$ If and only if $$\kappa(p) = \kappa +O(-p\log{p}) \,\,\ {\rm as} \,\, p \to 0,$$ then $$c_n = \exp(\kappa n +O(\log{n})).$$ We have generated data for this situation with a simple backtracking algorithm that gave data for walks up to length 24 steps. One could clearly do significantly better with more effort using more efficient algorithms based on  transfer matrices and finite lattice methods, but as we show below, even these short series are sufficient to provide strong support for the conjectured behaviour.
We remark that this result holds true even if the two layers are coupled along just a single line of sites. That is to say, one does not require all the orthogonal edges to be present, but only those along a single line of the lattice.

In fact Hammersley's model only requires the walk segments to be self-avoiding each time a layer is visited. That is to say, a walk returning to layer 1 from layer 2 need pay no heed to the preceding walk segment(s) on layer 1. The enumerations we have carried out are therefore for a proper subset of the Hammersley model.

Hammersely published a generalisation of this problem in 1991 in \cite{H91}. The proof is somewhat delicate, as evidenced by the fact that the following year Hammersley published a Corrigendum \cite{H92}, in which an error in the original proof was repaired.

From our 24 term data we generated series for $p=0.02,\,\,p=0.05,\,\,p=0.1,\,\,p=0.15$ and found $\exp(\kappa(p))=2.694,\,\,2.746,\,\,2.800,\,\,2.865$ respectively. Of course, when $p=0$ we have $\exp(\kappa)=\mu=2.63815853032790(3)$ \cite{JSG16}.

\begin{figure}[htbp]
   \centering
   \includegraphics[width=4.3in]{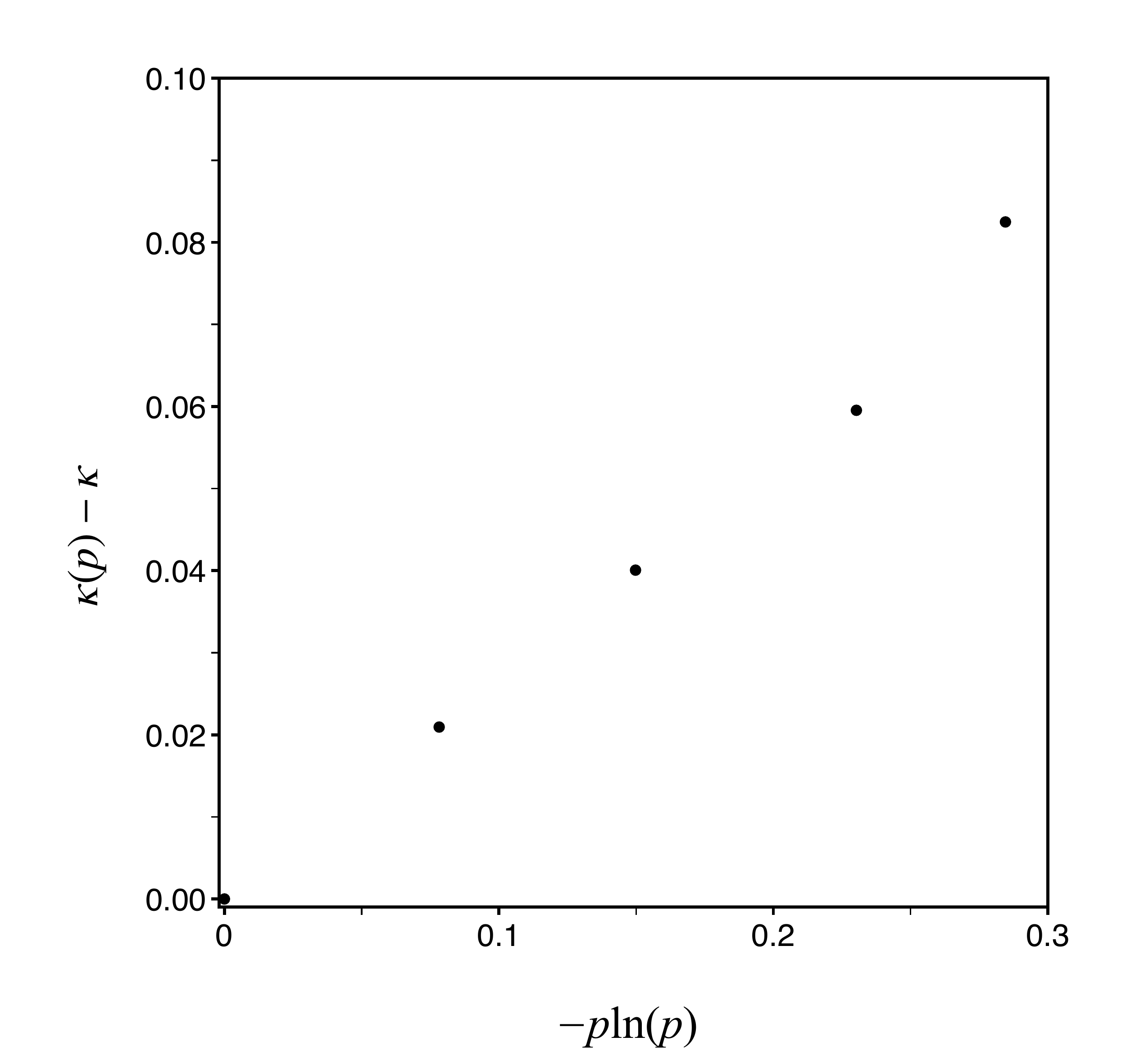} 
 \caption{Plot of $\kappa(p)-\kappa$ against $-p\log(p)$}
   \label{fig:kappa}
\end{figure}

We show in Figure~\ref{fig:kappa} a plot of $\kappa(p)-\kappa$ against $-p\log(p).$ This is seen to be visually linear, and we estimate the gradient to be around 0.27. That is to say, $\kappa(p)-\kappa \sim C p\log(p),$ where $C \approx -0.27.$ So the numerical evidence for the conjectured behaviour is seen to be quite strong.

\section{Coefficient ratio of ratios}

If $$r_n = \frac{c_{n+2}c_{n-2}}{c_n^2} = 1 + O \left ( \frac{1}{n^2} \right),$$ it follows that $$c_n = \exp(\kappa n + O(\log {n} )).$$ 
Alternatively, note that if $c_n \sim C\mu^n \cdot n^g,$ then $\frac{c_{n+2}c_{n-2}}{c_n^2} \sim 1 - \frac{4g}{n^2} .$ That is, this condition is both necessary and sufficient.

We remark that other ratios, such as $\frac{c_{n+1}c_{n-1}}{c_n^2}$ are expected to behave similarly, but there are at least two reasons for the choice made in the opening paragraph. Firstly, for loose-packed lattices such as the square lattice, there is a singularity of the generating function at $-1/\mu,$ and this introduces additional terms in the asymptotics of the ratios $c_n/c_{n-1}$ which are not present in the ratios $c_n/c_{n-2}.$ Secondly, the results proved for the ratios (discussed below) are for $c_n/c_{n-2}.$

\begin{figure}[htbp] 
   \centering
   \includegraphics[width=13cm]{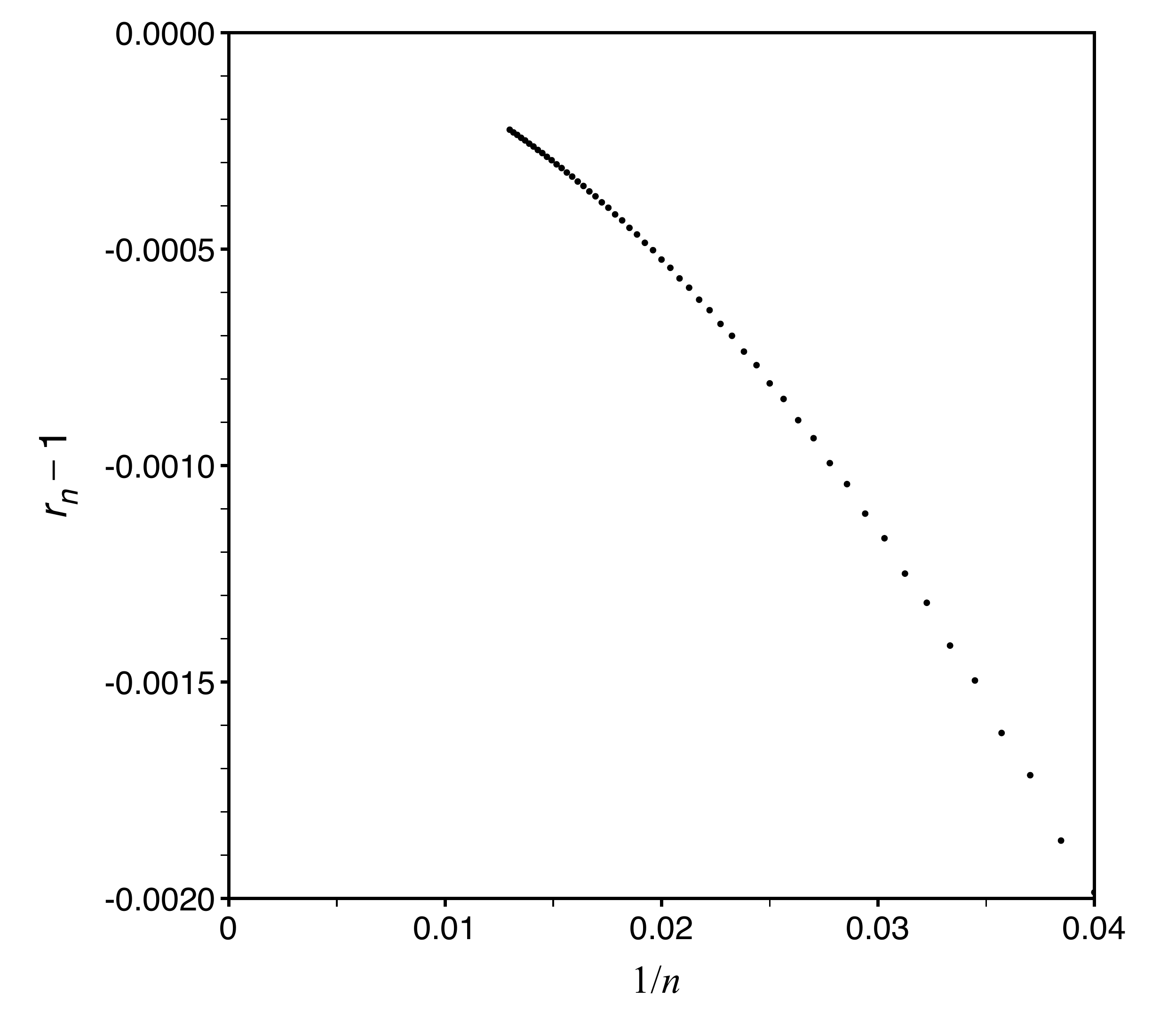} 
 \caption{Plot of $r_n-1 = \frac{c_{n+2}c_{n-2}}{c_n^2} - 1 $ against $1/n.$}
   \label{fig:rn-n}
\end{figure}

\begin{figure}[htbp] 
   \centering
   \includegraphics[width=13cm]{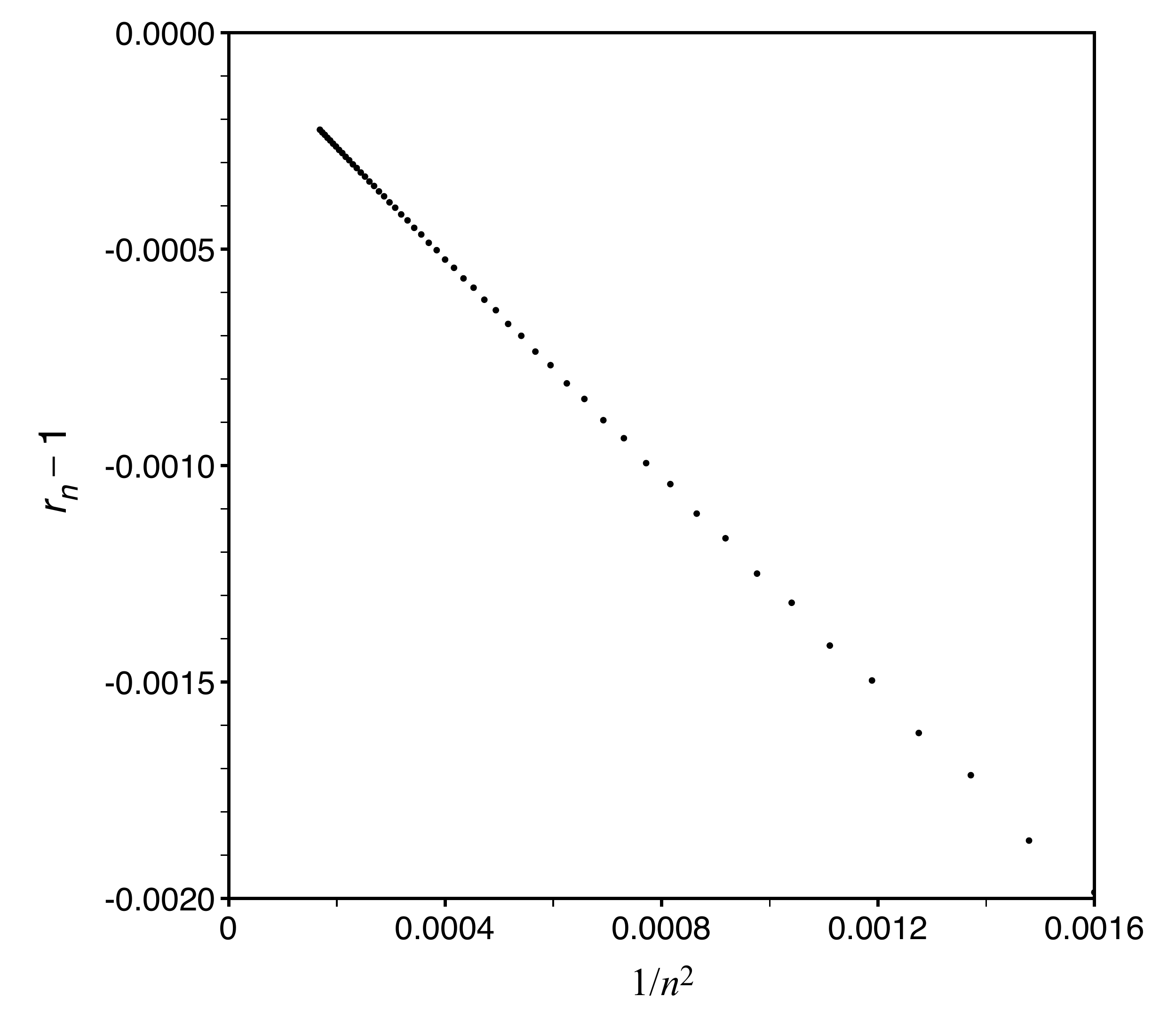} 
 \caption{Plot of $r_n-1 = \frac{c_{n+2}c_{n-2}}{c_n^2} - 1 $ against $1/n^2.$}
   \label{fig:rn-nsq}
\end{figure}

We have used the first 80 terms of the known series for square-lattice SAWs and plotted $r_n-1$ first against $1/n$ in Figure~\ref{fig:rn-n} and then against $1/n^2$ in Figure~\ref{fig:rn-nsq}. By construction, both plots must pass through the origin. There is considerable curvature in the first plot, which curvature must increase in order to pass through the origin, while the second plot is visually linear. 

This provides abundant numerical support for the conjectured behaviour. Unfortunately, we have been unable to prove the $1/n^2$ dependence. In fact what has been proved is a long way from this result. In \cite{K63} Kesten proved for SAWs that $$\lim_{n \to \infty} \frac{c_{n+2}}{c_n} = \mu^2,$$ while for SAPs the corresponding result is that $$\lim_{n \to \infty} \frac{p_{2n+2}}{p_{2n}} = \mu^2,$$ and for bridges $$\lim_{n \to \infty} \frac{b_{n+1}}{b_n} = \mu.$$ A nice proof of these results can be found in the book of Madras and Slade \cite{MS93}. As for the convergence rate, Kesten also proved in \cite{K63} that $$ \left | \frac{c_{n+2}}{c_n} - \mu^2 \right | \le Kn^{-1/3},$$ which is rather far from the result we need.

\section{Conclusion}
We have given three distinct approaches to improving the known asymptotics of two-dimensional SAWs, which are essentially due to J M Hammersley. We have provided abundant numerical evidence  that they are indeed based on assumptions that are true. Another observation of Hammersley is that ``Once you know something is true, it is easier to prove it.'' We hope that our calculations provide the necessary impetus for such a proof.

\section{Resources}
\label{sec:resource}

The enumeration data used in this study can be found at our GitHub repository \url{https://github.com/IwanJensen/Self-avoiding-walks-and-polygons/} by following the relevant branch.

\section*{Acknowledgements}
We would like to thank Geoffrey Grimmett, Neal Madras and Stuart Whittington for comments on an earlier draft of the manuscript, and Nathan Clisby, Alan Hammond and Neal Madras for their comments on a later draft. They are not responsible for its shortcomings. AJG would like to acknowledge financial support from ACEMS, the Australia Research Council Centre of Excellence for Mathematical and Statistical Frontiers.


\begin{thebibliography}{10}
\bibitem{BBS15} R Bauerschmidt, D C Brydges, G Slade {\em Logarithmic Correction for the Susceptibility of the 4-Dimensional Weakly Self-Avoiding Walk: A Renormalisation Group Analysis}, Commun. Math. Phys. {\bf 337 } (2015) 817--877.
\bibitem{BSTW17} R Bauerschmidt, G Slade, A Tomberg and B C Wallace {\em Finite-Order Correlation Length for Four-Dimensional Weakly Self-Avoiding Walk and $|\phi|^4$ Spins,} Ann. Henri Poincar\'e {\bf 18} (2017) 375--402.
\bibitem{C10} N. Clisby, {\em Accurate estimate of the critical exponent ν for self-avoiding walks via a fast
implementation of the pivot algorithm}, Phys. Rev. Lett. {\bf 104} (2010), 05570
\bibitem{C17} N. Clisby, {\em Scale-free Monte Carlo method for calculating the critical exponent $\gamma$ of self-avoiding walks},  J. Phys. A: Math. Theor. {\bf 50} (2017) 264003.
\bibitem{D86} B Duplantier, {\em Polymer network of fixed topology: renormalization exact critical exponent $\gamma$ in two dimensions, and $d=4-\epsilon$},  Phys. Rev. Lett. {\bf 57} (1986) 941--4.
\bibitem{DG19} B. Duplantier and A J Guttmann, {\em New scaling laws for self-avoiding walks: bridges and worms}, J. Stat. Mech:~Theor. and Exp. (2019) 104010.
\bibitem{DCH13} H Duminil-Copin and A Hammond, {\em Self-avoiding walk is sub-ballistic}, Comm. Math. Phys. {\bf 324} (2013) 401--23.
\bibitem{DCea18} H Duminil-Copin, S Ganguly, A Hammond and I Manolescu, {\em Bounding the number of self-avoiding walks: Hammersley-Welsh with polygon insertion,} Ann. Probab. {\bf 48} (2020) 1644--1692.
\bibitem{DGHM16} H Duminil-Copin, A Glazman, A Hammond and I Manolescu, {\em On the probability that self-avoiding walk ends at a given point}, Ann. Probab. {\bf 44} (2) (2016) 955--983. 
\bibitem{F23} M Fekete, {\em \"Uber die Verteilung der Wurzeln bei gewissen algebraischen Gleichungen mit ganzzahligen Koeffizienten}, Math. Zeit. {\bf 17} (1) (1923) 228--249. 
\bibitem{F49} P J Flory, {\em The configuration of real polymer chains}, J. Chem. Phys. {\bf 17} (1949), 303--310.
\bibitem{H61} J M Hammersley, {\em The number of polygons on a lattice}, Proc. Camb. Phil. Soc. {\bf 57} (1961) 516--523.
\bibitem{H91} J M Hammersley, {\em Self-avoiding walks}, Physica A {\bf 177} (1991) 51--57.
\bibitem{H92} J M Hammersley, {\em Corrigendum Self-avoiding walks}, Physica A {\bf 183} (1991) 574--578.
\bibitem{HM54} J M Hammersley and K W Morton, {\em Poor man's Monte Carlo,} J. Royal. Stat. Soc. Series B {\bf 16} (1954) 23--38.
\bibitem{HW62} J M Hammersley and D J A Welsh, { \em Further results on the rate of convergence to the connective constant of the hypercubical lattice}, Quart. J. Math. Oxford (2) {\bf 13} (1962) 108--110. 
\bibitem{H18} A Hammond, {\em An upper bound on the number of self-avoiding polygons via joining}, Ann. Probab. {\bf 46} (1) (2018) 175--206.
\bibitem{H19} A Hammond, {\em On self-avoiding polygons and walks: the snake method via pattern fluctuation}, Trans. Amer. Math. Soc. {\bf 372} (2019) 2335--2356.
\bibitem{H19a} A Hammond, {\em On self-avoiding polygons and walks: the snake method via polygon joining}, Electron. J. Probab. {\bf 24} article 49, 43pp (2019).
\bibitem{HS92} T. Hara and G. Slade, {\em Self-avoiding walk in five or more dimensions. I. The critical behaviour},
Commun. Math. Phys. {\bf 147} (1992) 101--136.
\bibitem{JSG16} J L Jacobsen, C R Scullard and A J Guttmann, {\em On the growth-constant for square-lattice self-avoiding walks}, J Phys A: Math. Theor. {\bf 49} (2016) 494004 (18pp).
\bibitem{J16} I Jensen, {\em Square lattice self-avoiding walks and biased differential approximants}, J Phys A: Math. Theor. {\bf 49} (2016) 424003 (13pp).
\bibitem{K63} H Kesten, {On the number of self-avoiding walks}, J. Math. Phys. {\bf 4} (1963) 960–-969.
\bibitem{LSW02} G Lawler, O Schramm and W Werner, {\em On the scaling limit of planar self-avoiding walk}, Fractal Geometry and Applications: A Jubilee of Benoit Mandelbrot, Proc. Sympos. Pure Math. vol. 72 part 2,  (2002), Amer. Math. Soc.,  Providence, RI. Also arXiv:math/0204277 
\bibitem{NM95} Neal Madras, {\em A rigorous bound on the critical exponent for the number of lattice trees, animals, and polygons}, J Stat Phys {\bf 78} (1995) 681--699. 
\bibitem{M14} N Madras, {\em A lower bound for the end-to-end distance of self-avoiding walk}, Can. Math. Bull. {\bf 57} (2014) 113--118.
\bibitem{MS93} N Madras and G Slade, {\em The self-avoiding walk}, (1993), Birkh\'auser, Boston.
\bibitem{N82} B Nienhuis, {\em Exact Critical Point and Critical Exponents of O$(n)$
Models in Two Dimensions}, Phys. Rev. Lett. {\bf 49} (1982) 1062--5.
\bibitem{O47} W J C Orr, {\em Statistical treatment of polymer solutions at infinite dilution}, Trans. Farady Soc. {\bf 43} (1947) 12--27.

\end{thebibliography}
\end{document}